\documentclass[pra,twocolumn,nofootinbib,preprintnumbers,superscriptaddress]{revtex4-2}

\usepackage{natbib}
\usepackage{amssymb,amsbsy,amsmath,amsfonts}
\usepackage{mathrsfs}
\usepackage{graphicx}
\usepackage{epsf,epsfig,latexsym,amsthm,fancyhdr,rotating}
\usepackage{graphics,psfrag,longtable}
\usepackage{slashed}
\usepackage{esint}
\usepackage{nicefrac}
\usepackage{braket}
\usepackage{mathtools}
\usepackage{ulem}

\newcommand{\nn}{\nonumber}
 
\renewcommand{\Re}{{\rm Re\,}}

\newcommand{\be}{\begin{equation}}
\newcommand{\ee}{\end{equation}}

\usepackage{color}
\definecolor{purple}{rgb}{.36,.12,.60}
\definecolor{orange}{rgb}{.9,.3,.0}
\definecolor{green}{rgb}{0,.6,0}
\definecolor{RED}{rgb}{.9,.3,.1}

\begin{document}

\title{Non-Diffractive Topological Spin Textures of Relativistic Twisted Fermion Beams}



\author{Andrei Afanasev}
\affiliation{Department of Physics,
The George Washington University, Washington, DC 20052, USA}
\author{Carl E. Carlson}
\affiliation{Physics Department, William \& Mary, Williamsburg, Virginia 23187, USA}

\begin{abstract}
In optics and acoustics, in structured beams, non-diffracting polarization measures in clearly diffracting beams, and spin direction distributions in the core of these waves that have Skyrmionic behavior have been found.  We here study the equivalents of these for fermions and show that in corresponding circumstances the non-diffractive spin textures persist independently of spin, statistics, or kinematics (or propagation speed of the structured wave).  As a part of the study, we present LG solutions for Dirac particles valid for both relativistic and non-relativistic kinematics.
\end{abstract}
\date{\today
}
\maketitle


\section{Introduction}


Topological polarization structures in propagating structured waves of light \cite{bliokh_rpp_2019} and sound \cite{muelas_prl_2022} are a subject of active research due to their impact on imaging applications, information transfer and storage. Of recent interest for structured vortex beams is the discovery of polarization features that maintain a fixed spatial size despite the beams spreading due to diffraction \cite{afanasev_advphot_nexus_2023}.  A number of such features were recently observed experimentally, and of even more recent particular interest is the distribution of longitudinal and transverse spin densities in a plane perpendicular to propagation direction of the overall wave~\cite{Annenkova:2025heg, mata2025skyrmionic, mata2026nonspread}.  These spin density distributions, or spin textures, can be mapped onto a spherical surface, or Poincaré sphere, giving patterns and winding numbers reminiscent of Skyrme models for nucleons the far past of nuclear and particle physics \cite{Skyrme:1962vh, Zahed:1986qz}.  


So far, the polarization features and spin textures just mentioned have been studied for bosons, including both acoustic waves and electromagnetic fields.  The results are as already stated: there are non-diffracting polarization features in an overall diffracting wave, and there are spin density distributions that map onto a Poincaré sphere with non-zero winding number.

Motivated by possible applications of vortex electron and neutron beams \cite{mcmorran2017origins,sarenac2018methods, sarenac2022experimental}, we will proceed to see if these features persist for fermion, here specifically spin-1/2, vortex fields.  We will look both at Laguerre-Gauss (LG) solutions and at Bessel waves.

We shall indeed show that the conditions that allow polarization features that maintain constant spatial dimension despite the overall spreading of the beam do exist in the fermionic case, with results analogous to the bosonic case, providing the corresponding conditions persist.

LG solutions are intrinsically diffracting solutions, so that propagation-independent non-diffracting substructures are of potentially great interest.   The LG solutions are valid in the paraxial approximation.  They have a Gaussian falloff away from the vortex axis, and are considered realistic as long as the conditions of the approximation are realized.  These conditions are basically that the Fourier representation of the stated is dominated by components whose wave number vectors or momentum vectors are at small angles to the direction of overall propagation of the state (see, $e.g.$ Ref.\cite{andrews2012angular}).

Existing literature and reviews, to our knowledge, give LG solutions that are valid non-relativistically and for the relativistic kinematics only show the Bessel solutions~\cite{Bliokh2017theory,afanasev2026vorticity}.  Hence we derive here LG solutions that are valid for all kinematics, both non-relativistic and relativistic.

The Bessel vortex solutions for fermions are well known and work well both relativistically and non-relativistically.  They are of course are non-diffracting.  Hence any studies showing non-diffractive polarization features are superfluous. However, the Skyrmion-like features of the spin distribution are still remarkable here, so the Bessel waves provide a continuation of this part of the study.
  
The LG solution for relativistic fermions will be displayed in an Appendix.  In the main text, Sec. II will show the non-diffracting polarization measure and the Skyrmionic structure for LG fermion solutions, Sec. III will comment on the Skyrme-like mapping for the Bessel solutions, and closing comments will be offered in Sec. IV.


\section{Spin mappings and Skyrmions for Laguerre-Gauss Fermion States}


In studying a LG or Bessel or other wave front for a field with spin, we become interested not only in the density distribution across the wave front, but also in the magnitude and orientation of the spind distribution across the wavefront.  For a spin-1/2 position dependent wave function, the spin vector at a given point is evaluated from 
\be
\vec S = \psi^\dagger \frac{1}{2} \vec\Sigma \psi   \,,
\ee
where $\vec\Sigma$ above is the $4 \times 4$ block diagonal matrix made from the standard $2 \times 2$ Pauli matrices. 

For a LG spin-1/2 field propagating in the $z$-direction and written in cylindrical coordinates 
$\{ \rho,\phi,z \}$
\begin{widetext}
\be                 \label{eq:lganti}
\psi = 	\left(	\begin{array}{c}
			a (E+m) f_\ell   \ \ \ 	\\[0.5 ex]
			b (E+m) f_{\ell+1}		\\[0.5 ex]
	a \left( k  \displaystyle{ - \frac{ \ell +1 }{ \zeta } + \frac{k \rho^2 }{2 \zeta^2 }  }  \right)	
		 f_\ell	
	+ i b \left( \displaystyle \frac{ k \rho^2 }{ 2 \zeta^2} - \frac{ 2(\ell+1) }{ \zeta }  \right) 
	   	f_{\ell}										\\[1 ex]
 i a \displaystyle	k	f_{\ell+1}
	- b \left( k \displaystyle{  - \frac{ \ell+2 }{ \zeta } + \frac{k \rho^2 }{2 \zeta^2 }  } \right)  
			f_{\ell+1} 
			\end{array}		\right)	.
\ee
\end{widetext}

This wave function is an eigenstate of $J_z$, the total angular momentum along the $z$-direction with eigenvalue $j_z = \ell +1/2$, and we have written it for the case $\ell \geq 0$ and restricted to polynomial quantum number $p=0$ for any Laguerre polynomials.  The functions containing the Gaussian falloff, monomial $\rho$ dependence, and azimuthal $\phi$ dependence are
\be
f_\ell = \frac{z_R}{\zeta} 
\left( \frac{\rho}{\zeta} \right)^{|\ell|}
\exp\left[ - \frac{ k \rho^2 }{ 2\zeta } 
    + i\ell\phi + i k z     \right] ,
\ee
where $\zeta = z_R + iz$ and $z_R$ is the Raleigh length.  More details are in the Appendix.

The coefficients $a$ and $b$ can be chosen arbitrarily.  We will speak of the pure $a$ term ($a \ne 0, b = 0$) as the parallel configuration since in the upper component the orbital and spin angular momentum point in the same direction for $\ell > 0$, and the pure $b$ term ($b \ne 0, a = 0$) will be the antiparallel configuration.

If we generically write the state as
\be         \label{eq:genstate}
\psi =  \left(
        \begin{array}{c}
        A \\ B \\ C \\ D
        \end{array}
        \right)     ,
\ee
then the longitudinal component of the spin, squared, will be
\be
| S_z |^2 = \frac{1}{4} \left(
|A|^2 - |B|^2 + |C|^2 - |D|^2 \right)^2    .
\ee 
The transverse spin squared will be
\begin{align}
| S_\perp |^2 &= | S_x |^2 + |S_y |^2 
            \nn\\
&=  |A|^2 |B|^2 + |C|^2 |D|^2 + 
2 \Re \left( A B^* C^* D \right)
\end{align}
and the total spin squared will be
\begin{align}
&| \vec S |^2 = |S_\perp|^2 + |S_z|^2 \nn\\
&\ \ = \frac{1}{4} \left(
|A|^2 + |B|^2 + |C|^2 + |D|^2 \right)^2
    -   | AD - BC |^2   .
\end{align}

A spin dependent quantity found to be of interest in the bosonic case~\cite{Annenkova:2025heg} is the spin alignment parameter, which measures the distribution of spin between the longitudinal and transverse directions,
\be
\chi_{\text{TL}} = 
\frac   { |S_\perp|^2 - |S_z|^2 }
        { |S_\perp|^2 + |S_z|^2 }
\ee

\begin{figure}[t!]
\includegraphics[width=1.0\columnwidth]
{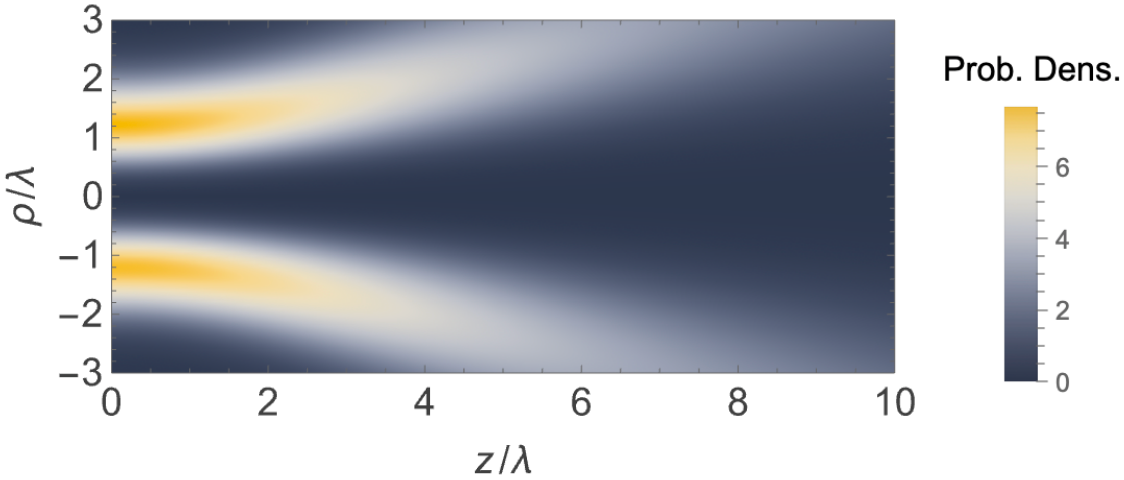}

\bigskip
\includegraphics[width=1.0\columnwidth] 
  {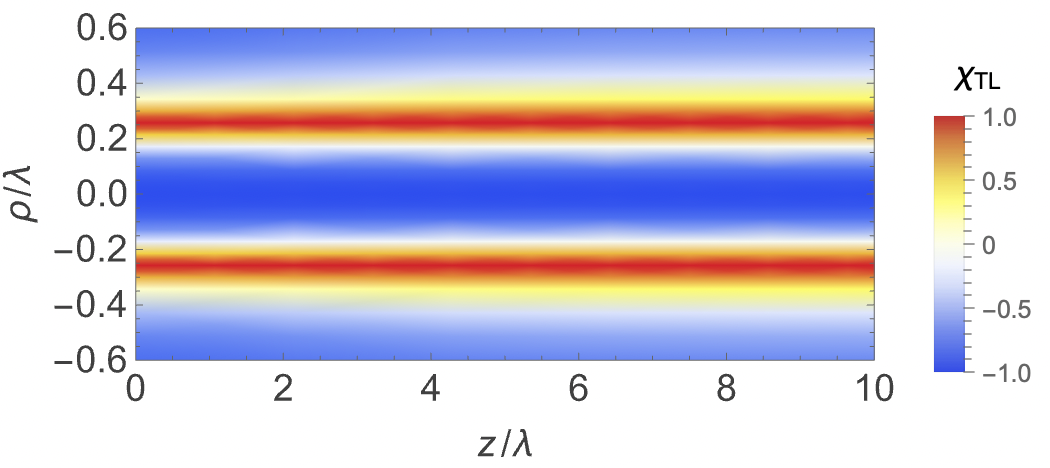}

\caption{Upper panel: Probability density for a LG wave, showing the wave clearly spreading;  Lower panel: the spin alignment parameter $\chi_\text{TL}$.   Both plots are for an antiparallel case with $\ell = 1$, $w_0 = \lambda\equiv2\pi/k$, and for fairly relativistic fermions, specifically Lorentz parameter $\gamma = 1.5$ or $\beta=v/c \approx 0.75$. The units for the probability density are arbitrary, the numbers for $\chi_\text{TL}$ are the actual numbers.}
\label{fig:one}
\end{figure}

The result for $\chi_\text{TL}$ is particularly striking for the antiparallel case.  We can show plots directly, Fig.~\ref{fig:one}, first of the probability density across the beam on the vertical axis and as a function of propagation distance $z$ along the horizontal axis, and then a corresponding plot of $\chi_\text{TL}$ for an antiparallel case.  For antiparallel case algebraically, the main terms contributing to $\chi_\text{TL}$ have the same $\zeta$ dependence and hence any $z$-dependence cancels in the ratio that give $\chi_\text{TL}$.  The result is an alignment parameter that, at least in the core of the wave, meaning at small $\rho$, does not expand (or shrink) as the wave propagates.

Algebraically,  one can find for the antiparallel case, with the Compton wavelength $\lambdabar_C\equiv 1/m_e$ serving as the length scale,
\begin{align}
&\chi_\text{TL} = -1 +  \\
&\frac{ 8 (\gamma^2-1) (\ell+1)^2
                \lambdabar_C^2 \rho^2}
{ 4 (\ell +1)^4 \lambdabar_C^4 -
4(\gamma+1) (\ell+1)^2 \lambdabar_C^2 \rho^2
+ \gamma^2 (\gamma+1)^2 \rho^4 }  , \nn
\end{align}
where $\gamma$ is the Lorentz parameter 
$\gamma = (1-\beta^2)^{-1/2}
=(1-(v/c)^2)^{-1/2}$.  (Some small terms have been dropped in the above, though the plots are made with the full expressions, with no visible difference.) The result for $\chi_\text{TL}$ in the core for the Bessel beam case is just the same as given above, as we shall show in the next Section.

The parameter $\chi_\text{TL}$ peaks at
\be
\rho_\text{peak} = 
\sqrt{ \frac{2}{\gamma (\gamma+1) } } \ 
(\ell+1) \lambdabar_C \,,
\ee
and the peak value of $\chi_\text{TL}$ is always $+1$ (although the limit can be touchy for $\gamma = 1$, the result implies that the value of $\rho_{peak}$ is momentum-independent for the non-relativistic particles).

For completeness we should mention that the $z$-independence of $\chi_\text{TL}$ does not persist in the parallel case, where for $a \ne 0, b=0$ one can find
\be
\chi_\text{TL}^\text{parallel}
    = -1 + \frac{ (\gamma-1)^2}{\gamma^2}
    \frac{ \rho^2 }{ z_R^2+z^2 }
+ \mathcal O\left(
    \frac{ \rho^4 }{ (z_R^2+z^2)^2 } \right) .
\ee
The spreading of $\rho$ with increasing $z$ to achieve a fixed value of $\chi_\text{TL}$ is manifest.

The unit vector that gives the spin direction on a transverse plane at a fixed $z$-value is independent on $z$. For the antiparallel case, it is 
\begin{align}
\hat s &= (\text{norm})\bigg[
-2\sqrt{\gamma^2-1} (\ell+1) \lambdabar_C \rho \, \hat\phi     \nn\\       & \quad +
\left( 2(\ell+1)^2 \lambdabar_C^2 -
\gamma(1+\gamma) \rho^2 \right) \hat z
    \bigg]  \,,
\end{align}
where the (norm) is chosen to ensure $| \hat s | = 1$.   

\begin{figure}[t!]
\includegraphics[width=0.7\columnwidth] {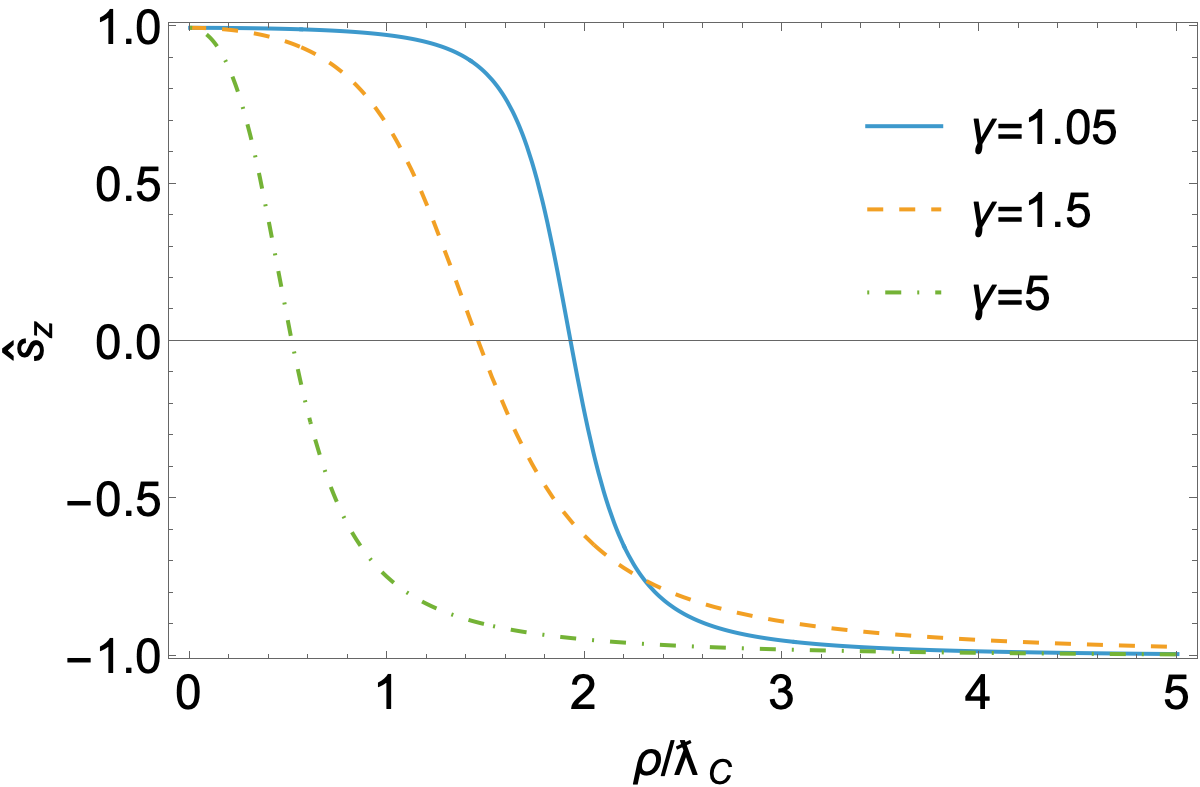}

\bigskip
\includegraphics[width=0.7\columnwidth] 
  {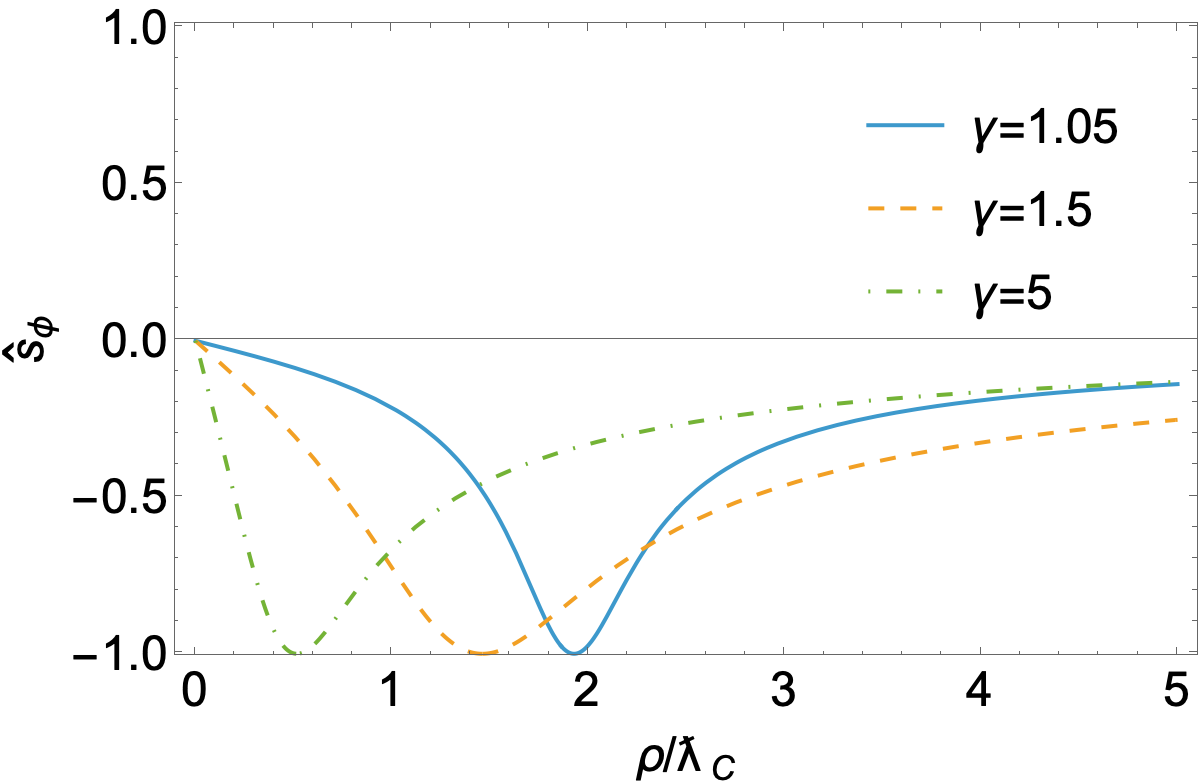}

  \bigskip
\includegraphics[width=0.7\columnwidth] 
  {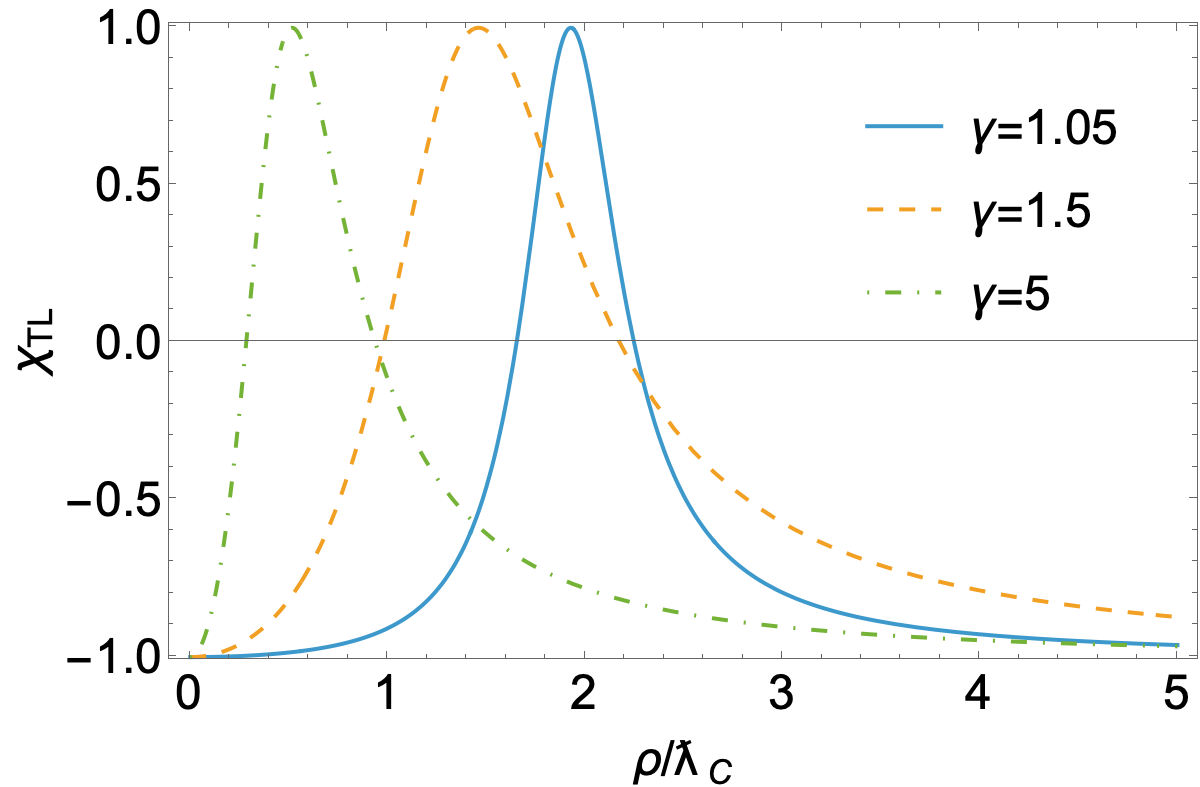}

\caption{Upper panel: A longitudinal component of (normalized) spin density ($\hat s_z$);  Middle panel:  An azimuthal component of the same ($\hat s_\phi$); Lower panel: 
The spin alignment parameter $\chi_\text{TL}$.   All plots are as a function of the radial distance to the vortex center (in units of reduced Compton wavelength $\lambdabar_C\equiv1/m_e$). Shown is an antiparallel case with $\ell = 1$,  and Lorentz parameter $\gamma = 1.05, 1.5$ and 5.0, respectively indicated by solid, dashed and dashed-dotted lines.}
\label{fig:two}
\end{figure}

Spin components and the alignment parameter $\chi_{TL}$ are presented in Fig.\ref{fig:two}. One can see that for non-relativistic particles, the quantity $\hat s_z$ changes sign near $\rho\approx 2\lambdabar_C$, at the same values where $\chi_{TL}$ reaches maximum. As the beam energy increases, the spin texture shrinks toward the vortex center.

\begin{figure}[t!]
\includegraphics[width=0.95\columnwidth] {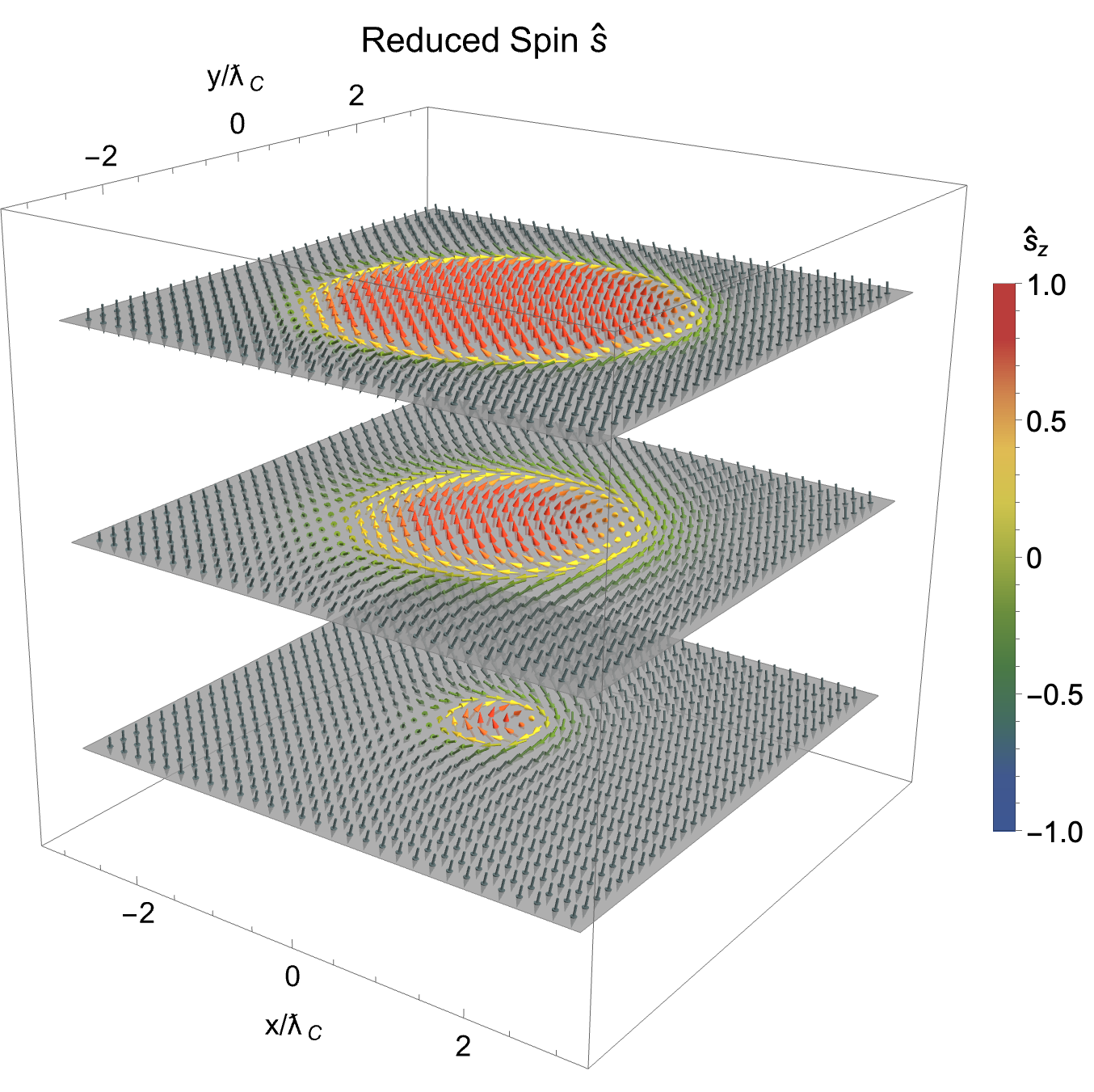}
\caption{
Three-dimensional representation of the normalized spin polarization vector $\hat s$ at a fixed value of $z$. The spin flips its sign with increasing $\rho$ while remaining perpendicular to the vector $\vec \rho$; this behavior indicates Bloch-type Skyrmion topology. Upper, middle, and lower plots correspond to Lorentz factor $\gamma = 1.05, 1.5$ and 5.0, respectively. The spin texture is independent of the propagation distance $z$.}
\label{fig:three}
\end{figure}

The vector field of $\hat s$ is represented in three dimensions in Fig.\ref{fig:three}. It is apparent that the region of positive $\hat s_z$-component shrinks toward small values of $\rho$
as the beam's momentum increases.

The Skyrme number can be calculated from
\begin{align}
   N_{SK} = \frac{1}{4\pi} \int \hat s \cdot\left( \frac{\partial \hat s}{\partial x} \times 
\frac{\partial \hat s}{\partial y} \right) dx dy     \,.
\end{align}
For a generic situation with cylindrical symmetry and no radial spin component,
\be
\hat s = f(\rho) \,\hat\phi + g(\rho)\,\hat z
\ee
with a normalization condition
\be
f(\rho)^2 + g(\rho)^2 = 1
\ee
and the associated relation
\be
f f'+g g'=0,
\ee
the integrals can be done analytically, and integrating with respect to $\rho$ from the center to a value $\rho_\text{max}$ yields
\be
N_{SK} = -\frac{1}{2}[g(\rho_{max})-g(0)].
\ee
It implies that if the longitudinal spin component dominates at $\rho=0$ ($g(0)=1$) and disappears at $\rho_{max}$ ($g(\rho_{max})=0$), then $N_{SK}=1/2$. If, on the other hand, the longitudinal spin component flips the sign ($g(\rho_{max})=-1$), then Skyrme number is unity, $N_{SK}=1$.

The fact that the transverse spin has only an azimuthal component allows us to classify this topological spin texture as a Bloch-type spin meron ($N_{SK}=1/2$) or Skyrmion ($N_{SK}=1$), $c.f.$ Ref.\cite{Annenkova:2025heg}. The latter is the case described by Eq.(12), with $\rho_{max}$ extended to large values.


\section{Bessel Beams}


Bessel states are non-diffracting, so finding non-spreading polarization metics is not news.  However, the spin texture on the wavefront of a Bessel beam and its projection onto the 
Bloch sphere is still of interest.  In particular, we would like to compare the Bessel beam spin texture in the core of the vortex to the corresponding result for the LG beam.

One can write Bessel beams where all the contributing plane wave states have the same helicity~\cite{Serbo:2015kia} or one can take linear combinations of these to allow spin states with varying spin properties~\cite{Bialynicki-Birula:2016unl,Barnett:2017wrr,afanasev2026vorticity}. A general expression is~\cite{afanasev2026vorticity}
\begin{align}               \label{eq:BB2}
\psi_{B} \propto
	\left(	\begin{array}{c}
		a (E+m_e)	\, f_B^{\ell}	\\[1ex]
		b (E+m_e)	\, f_B^{\ell+1}	\\[1ex]
		(a k_z -i b \kappa )	\, 
        f_B^{\ell}	\\[1ex]
		(ia \kappa - b k_z )	\, 
        f_B^{\ell+1}
				\end{array}		\right)	.
\end{align}
These are energy and $J_z$ eigenstates, with $j_z = \ell+1/2$, and
\be
f_B^\ell = \exp(ik_z z + i \ell \phi) 
    J_\ell(\kappa\rho)  \,.
\ee
All the contributing Fourier states have momenta or vector wave number at a polar angle or pitch angle $\theta_k$ to the $z$-axis and 
\be
\kappa = k \sin\theta_k \,, \quad
    k_z = k \cos\theta_k    \,.
\ee

Since we have looked at antiparallel states (defined by $b\ne 0, a = 0$) for the LG case, we shall continue with the same here, whence
\begin{align}
\psi_{B} \propto
	\left(	\begin{array}{c}
		0	\\[1ex]
		(E+m_e)	\, e^{i\phi}	\\[1ex]
		-i \kappa 	\, 
        J_\ell(\kappa\rho)/
        J_{\ell+1}(\kappa\rho)  \\[1ex]
		- k_z	\, e^{i\phi}
				\end{array}		\right)	.
\end{align}
Being interested in the behavior in the core, or at small $\rho$ we use the expansions of the Bessel functions to find
\be
\frac{J_\ell(\kappa\rho)}
        {J_{\ell+1}(\kappa\rho)}
\approx \frac{ 2(\ell+1)}{\kappa\rho}
    - \frac{\kappa\rho}{2(\ell+2)}  .
\ee
We kept the next to leading term so we can evaluate its importance.  In evaluating $|S_z|^2$ or the probability density, we encounter expressions like, recalling the notation in Eq.~\eqref{eq:genstate},
\be
T = \pm \left( |B|^2 + |D|^2 \right)+ |C|^2,
\ee
where a cross term in $|C|^2$ is order $\rho^0$, just like the $|B|^2$ and $|D|^2$ terms.  Rewriting,
\begin{align}
T \propto \frac{4(\ell+1)^2}{\rho^2}
&+ \left( \pm \left[ (E+m_e)^2 + k_z^2 \right]
 - \frac{2(\ell+1) \kappa^2}{\ell+2} \right)
    \nn\\
&+ \frac{ \rho^2 }{ 4(\ell+2)^2 }    \,.
\end{align}
when we make the paraxial approximation, meaning small $\theta_k$, given the existence of large comparison terms in the $\rho^0$ term, we can 
neglect the $\kappa^2\rho^0$ term.  We can also in the same approximation substitute $k$ for $k_z$.  Finally also dropping the $\rho^2$ in the core, we obtain the expression
\be
T \propto \frac{4(\ell+1)^2}{\rho^2}
\pm 2E (E+m_e)  \,,
\ee
that is free of $\theta_k$ dependence (which controls transverse-plane distribution of Bessel state's probability density) and is precisely the same as the corresponding LG result.

Stated differently, in the core for the antiparallel case in the paraxial approximation, we can work from
\begin{align}               \label{eq:anticore}
\psi_{B} \propto
	\left(	\begin{array}{c}
		0	\\[1ex]
		(E+m_e)	\, e^{i\phi}	\\[1ex]
		-2 i  (\ell+1) / \rho
          \\[1ex]
		- k	\, e^{i\phi}
				\end{array}		\right)	,
\end{align}
again the same as the LG case, and so the spin textures will be the same in the two cases.

Some details on the derivations are in order. In obtaining Eq.~\eqref{eq:anticore} for LG antiparallel states, approximations analogous to the ones used here are needed.  In particular, looking at the $b$ terms in Eq.~\eqref{eq:lganti}, in the fourth component, one can easily drop the $\rho^2$ in the core, and the $(\ell+2)/\zeta$ will be small enough to ignore as long a $z_R$ is large compared to the wavelength.  In the third component,  the $\rho^2$ term can be more worrisome, but with some manipulation it leads to a term of order $k^2 (\ell+1) \lambdabar/(2\pi z_R)$ compared to an $E^2$ term in $S_z^2$ or probability density, so it will be small for non-relativistic kinematics and still small for any kinematics as long as $z_R$ is significant and $\ell$ not large.


\section{Summary and Conclusions}


We have looked at polarization phenomena for twisted fermions, with an interest in seeing if the results known in optics and acoustics \cite{Annenkova:2025heg, mata2025skyrmionic, mata2026nonspread} are replicated here.  We have studied mainly LG versions of twisted fermion beams, and have used expressions that are valid at all kinematics, not just in the non-relativistic limit.  LG states represent real conditions where the state is diffracting, or spreading significantly as it propagates in vacuum.  

Interestingly enough, it was found that for a twisted fermion in a certain configuration that there were polarization measures, ratios of polarization components, that remained fixed constant at a given distance from the beam's vortex line even as the beam overall spreads.  Further, one can look at the distribution of spin direction on on a plane perpendicular to the overall propagation direction and map it onto a sphere in a standard way and discover that one has a configuration with a non-zero winding number, like Skyrmions studied in particle physics.  The twisted state that gives the non-diffracting polarization results in the core is the antiparallel one, wherein the largest orbital angular momentum contribution and the largest spin contribution point in opposite directions.

We find that the polarization and formation of topological spin textures are not spin dependent or statistics dependent or kinematics dependent.
It was shown to be due to a continuity requirement in both optics and acoustics \cite{afanasev_advphot_nexus_2023,Annenkova:2025heg,mata2025skyrmionic,mata2026nonspread}. Since Dirac equation was derived under condition of continuity for the probability current, we observe  that continuity is likely a common cause for such behavior for both fermions and bosons.  We have shown that the spin alignment parameter, a measure of the relative size of the transverse and longitudinal polarizations, is non-diffractive for core radii also in the fermion antiparallel LG state, and the phenomenon can be seen for non-relativistic kinematics as well as for states propagating near the speed on light.  The same can be said for the fermionic spin texture mapping into Skyrmion-like states.  Naturally, the twisted fermion of the Bessel state is non-diffractive, but the realization of Skyrmion states from the core spin distribution is the same as for the LG states. 

In summary, we identified non-diffractive spin textures around the vortex core of longitudinally polarized (diffractive) LG fermion beams; they coincide with the corresponding spin textures of Bessel beams in a paraxial limit.


\section*{Acknowledgements}

A.A.~thanks Army Research Office for support under grant W911NF-23-1-0085. C.E.C.~thanks the National Science Foundation (USA) for support under grant PHY-1812326.



\appendix

\section{General twisted fermion LG solutions}

 
The LG solutions are monoenergetic solutions that propagate in a definite direction, fall off as a Gaussian in directions transverse to the propagation direction, and spread as they propagate forward.  

The LG solution for a spin-1/2 field must satisfy the Dirac equation.  As is well known, this also means the individual components of the Dirac solution satisfy the Klein-Gordon equation, which allows finding solutions that satisfy the LG conditions for the individual components.

One can view the Dirac equation as determining the lower components of the solution from the upper components.  This allows freedom in choosing the upper components, as long as they satisfy the Klein-Gordon equation, or the LG approximation (i.e., mostly forward propagating Fourier component waves) to it.  One proceeds by choosing the upper components to have the same Laguerre properties as the scalar wave,  using the Dirac equation to find the lower components, and noting that because derivatives commute, the lower components will satisfy the LG equation if the upper components do.

Explicitly, the Dirac equation is
\begin{align}
( \slashed{k} - m_e ) \psi =
    ( \slashed{k} - m_e ) 
    \left(  \begin{array}{c}
                \xi  \\
                \chi
            \end{array}     \right) = 0,
\end{align}
where $\xi$ and $\chi$ are two-component spinors, and $m_e$ is the mass of the fermion, and we use units where $\hbar = c = 1$. One can relate $\chi$ to $\xi$ using
\be
\chi = \frac{ \vec\sigma\cdot\vec k}{E+m_e} \xi,
\ee
where we used the standard representation of the Dirac $\gamma$-matrices, the $\vec\sigma$ are the standard $2 \times 2$ Pauli matrices, and the $\vec k$ in coordinate space are represented by derivative operators.

For the upper components take
\be
\xi =   \left(    
        \begin{array}{l}
        a (E+m_e) f_p^\ell      \\[1 ex]
        b (E+m_e) f_p^{\ell+1} 
        \end{array}
        \right)     .
\ee
We have written the state so that the angular momentum projected in the $z$-direction, $j_z$, is the same for both components, with $j_z = \ell + 1/2$.  (This need not always be done, see for example~\cite{Barnett:2017wrr}.)  The $a$ and $b$ coefficients are arbitrary, the $E+m_e$ factor is inserted for convenience, and the scalar LG solutions in cylindrical coordinates $\rho,\phi,z$ are
\be
f_p^\ell = \frac{z_R}{\zeta} 
\left( \frac{\rho}{\zeta} \right)^{|\ell|}
L_p^\ell(x)
\exp\left[ - \frac{ k \rho^2 }{ 2\zeta } 
    + i\ell\phi + i k z     \right] ,
\ee
where $\zeta = z_R + iz$ and $L_p^\ell$ is the Laguerre polynomial, and
\be
x = \frac {k z_R \rho^2 }{ z_R^2 + z^2 } 
\equiv \frac{ 2\rho^2 }{ w(z)^2 } 	\,.
\ee

The solution has two parameters.  One is the energy $E$, for which one may substitute the momentum magnitude $k = \sqrt{ E^2 - m_e^2}$ or the de Broglie wavelength $\lambda = 2\pi/k$.    The other parameter used here is the Raleigh length $z_R$.  Alternatively, one may use the beam width parameter $w_0$, which may be obtained by expanding one term in the exponential,
\be
\exp\left[ - \frac{ k\rho^2}{2\zeta} \right] =
    \exp\left[ - \frac{ \rho^2 }{w^2(z)}
    + i \frac{ k \rho^2 z}{2(z_R^2+z^2)} ,
\right]
\ee 
with
\be
w^2(z) = w_0^2 ( 1 + z^2/z_R^2 ) \quad
\text{and}     \quad
w_0^2 = 2z_R/k  \,.
\ee 
I.e., $w_0$ controls the magnitude of the Gaussian falloff and $z_R$ controls the propagation lengths $z$ where the beam spread becomes noticeable.

The full state is
\be
\psi = 	\left(	\begin{array}{c}
				a (E+m) f_{p,\ell}   \ \ \ 		\\
				b (E+m) f_{p,\ell+1}		\\
				a k_z f_{p,\ell}	+ b e^{-i\phi} (  k_\rho -ik_\phi  )  f_{p,\ell+1}								\\
	a e^{i \phi} (  k_\rho + ik_\phi  )  f_{p,\ell} + b k_z f_{p,\ell+1}
			\end{array}		\right)	,
\ee
where, again, the $k_i$ are derivative operators.

Working out the indicated steps, we obtain the solution for general
$\ell$ and $p$. We can soon make some restrictions and simplifications.
\begin{widetext}
\be
\psi = 	\left(	\begin{array}{c}
				a (E+m) f_p^{\ell}   \ \ \ 		\\[1 ex]
				b (E+m) f_p^{\ell+1}		
													\\
				\left[ a k_z^{(\ell)}
	- i b  \displaystyle \frac{ 2( \ell+p+1) }{ \rho}	
		\left( \frac{ \rho }{ \zeta } \right)^{ \textrm{sgn}(\ell) }	\right]	f_p^\ell	
		- \displaystyle \frac{ 2i a z (p+\ell) }{ \zeta^2 }	f_{p-1}^\ell		
			+ ib	\left[ \frac{ k \rho }{ \zeta } 	+ \frac{ \ell+1 - | \ell+1 | }{\rho}	\right]
				e^{-i \phi} f_{p}^{\ell+1}					
													\\
	 i a \displaystyle \frac{k\rho}{\zeta}  \left( \frac{ \zeta }{ \rho } \right)^{ \textrm{sgm}(\ell) }
	 	 \Big[ f_{p,}^{\ell+1} + f_{p-1}^{\ell+1}  \Big]
	+ i a e^{i\phi} \frac{ \ell - | \ell | }{ \rho } f_p^\ell
	 	- b \left[  k_z^{(\ell+1) }   f_{p}^{\ell+1}
			- i \displaystyle \frac{ 2z (p + \ell + 1 ) }{ \zeta^2 }	f_{p-1}^{\ell+1} 	 \right]	
			\end{array}		\right)	.
\ee
\end{widetext}
We have used the compact notation
\be
k_z^{(\ell)} =  k  \displaystyle{  - \frac{ | \ell | + 2p +1 }{ \zeta } 
					+ \frac{k \rho^2 }{2 \zeta^2 }  } 	  .
\ee
The quantity $\textrm{sgn}(\ell)$ is $1$ for $\ell$ non-negative and $-1$ for $\ell$ negative.  When $p=0$ the above result applies with the protocol $f_{-1}^\ell = 0$.

We simplify the result first by restricting it to $\ell \geq 0$, and then further restricting it to $p = 0$.  Recall that $L_0^\ell = 1$ for all $\ell$.  The LG Dirac solution, using $f_\ell = f_0^\ell$, becomes
\begin{widetext}
\be
\psi = 	\left(	\begin{array}{c}
			a (E+m) f_\ell   \ \ \ 	\\[0.5 ex]
			b (E+m) f_{\ell+1}		\\[0.5 ex]
	a \left( k  \displaystyle{ - \frac{ \ell +1 }{ \zeta } + \frac{k \rho^2 }{2 \zeta^2 }  }  \right)	
		 f_\ell	
	+ i b \left( \displaystyle \frac{ k \rho^2 }{ 2 \zeta^2} - \frac{ 2(\ell+1) }{ \zeta }  \right) 
	   	f_{\ell}										\\[1 ex]
 i a \displaystyle	k	f_{\ell+1}
	- b \left( k \displaystyle{  - \frac{ \ell+2 }{ \zeta } + \frac{k \rho^2 }{2 \zeta^2 }  } \right)  
			f_{\ell+1} 
			\end{array}		\right)	.
\ee
\end{widetext}

An alternative derivation of the fermion LG state is to use the Foldy-Wouthuysen transformation~\cite{Foldy:1949wa}, which for free Dirac particles can be both implemented and inverted analytically, and was used in a related twisted fermion context in~\cite{Barnett:2017wrr}).

The Foldy-Wouthuysen transformation decouples the upper and lower components of the Dirac spinor, and only the upper components are nonzero for positive energy solutions.  These upper components have no constraints beyond satisfying the Klein-Gordon equation, and a LG solution for each of the components can immediately be given (with, incidentally, the option that they have different $j_z$~\cite{Barnett:2017wrr}). Transforming back to the usual Dirac representation (for the same $j_z$ case) gives precisely the results quoted earlier in this Appendix.


\bibliography{vortfermi}

@article{Annenkova:2025heg,
    author = "Annenkova, Elena and Afanasev, Andrei and Brasselet, Etienne",
    title = "{Universal nondiffractive topological spin textures in vortex cores of light and sound}",
     journal = "arXiv preprint",
    eprint = "2512.02964",
    archivePrefix = "arXiv",
    primaryClass = "physics.optics",
    month = "12",
    year = "2025"
}

@article{Barnett:2017wrr,
    author = "Barnett, Stephen M.",
    title = "{Relativistic Electron Vortices}",
    doi = "10.1103/PhysRevLett.118.114802",
    journal = "Phys. Rev. Lett.",
    volume = "118",
    number = "11",
    pages = "114802",
    year = "2017"
}

@article{Bialynicki-Birula:2016unl,
    author = "Bialynicki-Birula, Iwo and Bialynicka-Birula, Zofia",
    title = "{Relativistic Electron Wave Packets Carrying Angular Momentum}",
    eprint = "1611.04445",
    archivePrefix = "arXiv",
    primaryClass = "quant-ph",
    doi = "10.1103/PhysRevLett.118.114801",
    journal = "Phys. Rev. Lett.",
    volume = "118",
    number = "11",
    pages = "114801",
    year = "2017"
}

@article{Foldy:1949wa,
    author = "Foldy, Leslie L. and Wouthuysen, Siegfried A.",
    title = "{On the Dirac theory of spin 1/2 particle and its nonrelativistic limit}",
    doi = "10.1103/PhysRev.78.29",
    journal = "Phys. Rev.",
    volume = "78",
    pages = "29--36",
    year = "1950"
}

@article{Serbo:2015kia,
    author = "Serbo, V. and Ivanov, I. P. and Fritzsche, S. and Seipt, D. and Surzhykov, A.",
    title = "{Scattering of twisted relativistic electrons by atoms}",
    eprint = "1505.02587",
    archivePrefix = "arXiv",
    primaryClass = "physics.atom-ph",
    doi = "10.1103/PhysRevA.92.012705",
    journal = "Phys. Rev. A",
    volume = "92",
    number = "1",
    pages = "012705",
    year = "2015"
}

@article{Skyrme:1962vh,
    author = "Skyrme, T. H. R.",
    title = "{A Unified Field Theory of Mesons and Baryons}",
    doi = "10.1016/0029-5582(62)90775-7",
    journal = "Nucl. Phys.",
    volume = "31",
    pages = "556--569",
    year = "1962"
}

@article{Zahed:1986qz,
    author = "Zahed, I. and Brown, G. E.",
    title = "{The Skyrme Model}",
    reportNumber = "PRINT-86-0160 (STONY-BROOK)",
    doi = "10.1016/0370-1573(86)90142-0",
    journal = "Phys. Rept.",
    volume = "142",
    pages = "1--102",
    year = "1986"
}

@article{mata2025skyrmionic,
  title={Skyrmionic polarization texture around the phase singularity of optical vortices},
  author={Mata-Cervera, Nilo and Sharma, Deepak K and Shen, Yijie and Paniagua-Dominguez, Ramon and Porras, Miguel A},
  journal={Physical Review Letters},
  volume={135},
  number={3},
  pages={033805},
  year={2025},
  publisher={APS}
}

@article{mata2026nonspread,
	author={Mata Cervera, Nilo and Porras, Miguel Angel and Shen, Yijie},
	title={Non-spreading meronic spin defects around optical vortices},
	journal={Reports on Progress in Physics},
	url={http://iopscience.iop.org/article/10.1088/1361-6633/ae7cb0},
    doi={10.1088/1361-6633/ae7cb0},
	year={2026}
}

@ARTICLE{Bliokh2017theory,
       author = {{Bliokh}, K.~Y. and {Ivanov}, I.~P. and {Guzzinati}, G. and {Clark}, L. and {Van Boxem}, R. and {B{\'e}ch{\'e}}, A. and {Juchtmans}, R. and {Alonso}, M.~A. and {Schattschneider}, P. and {Nori}, F. and {Verbeeck}, J.},
        title = "{Theory and applications of free-electron vortex states}",
      journal = {Phys. Rep.},
         year = {2017},
        month = may,
       volume = {690},
        pages = {1-70},
          doi = {10.1016/j.physrep.2017.05.006},
archivePrefix = {arXiv},
       eprint = {1703.06879},
 primaryClass = {quant-ph},
       adsurl = {https://ui.adsabs.harvard.edu/abs/2017PhR...690....1B}
}

@article{afanasev2026vorticity,
  title={Vorticity of Twisted Electron Fields: Role of the Energy--Momentum Tensor},
  author={Afanasev, Andrei and Carlson, Carl E and Mukherjee, Asmita},
  journal={Quantum Beam Science},
  volume={10},
  number={2},
  pages={8},
  year={2026},
  publisher={MDPI},
  DOI = {10.3390/qubs10020008}
}

@article{bliokh_rpp_2019,
  title={Geometric phases in 2D and 3D polarized fields: geometrical, dynamical, and topological aspects},
  author={Bliokh, Konstantin Y and Alonso, Miguel A and Dennis, Mark R},
  journal={Reports on Progress in Physics},
  volume={82},
  number={12},
  pages={122401},
  year={2019},
  publisher={IOP Publishing}
}

@article{muelas_prl_2022,
  title={Observation of polarization singularities and topological textures in sound waves},
  author={Muelas-Hurtado, Ruben D and Volke-Sep{\'u}lveda, Karen and Ealo, Joao L and Nori, Franco and Alonso, Miguel A and Bliokh, Konstantin Y and Brasselet, Etienne},
  journal={Physical Review Letters},
  volume={129},
  number={20},
  pages={204301},
  year={2022},
  publisher={APS}
}

@article{afanasev_advphot_nexus_2023,
  title={Nondiffractive three-dimensional polarization features of optical vortex beams},
  author={Afanasev, Andrei and Kingsley-Smith, Jack J and Rodr{\'\i}guez-Fortu{\~n}o, Francisco J and Zayats, Anatoly V},
  journal={Advanced Photonics Nexus},
  volume={2},
  number={2},
  pages={026001--026001},
  year={2023},
}

@book{andrews2012angular,
  title={The angular momentum of light},
  author={Andrews, David L and Babiker, Mohamed},
  year={2012},
  publisher={Cambridge University Press}
}

@article{mcmorran2017origins,
  title={Origins and demonstrations of electrons with orbital angular momentum},
  author={McMorran, Benjamin J and Agrawal, Amit and Ercius, Peter A and Grillo, Vincenzo and Herzing, Andrew A and Harvey, Tyler R and Linck, Martin and Pierce, Jordan S},
  journal={Philosophical transactions. Series A, Mathematical, physical, and engineering sciences},
  volume={375},
  number={2087},
  pages={20150434},
  year={2017}
}

@article{sarenac2022experimental,
  title={Experimental realization of neutron helical waves},
  author={Sarenac, Dusan and Henderson, Melissa E and Ekinci, Huseyin and Clark, Charles W and Cory, David G and DeBeer-Schmitt, Lisa and Huber, Michael G and Kapahi, Connor and Pushin, Dmitry A},
  journal={Science Advances},
  volume={8},
  number={46},
  pages={eadd2002},
  year={2022},
  publisher={American Association for the Advancement of Science}
}

@article{sarenac2018methods,
  title={Methods for preparation and detection of neutron spin-orbit states},
  author={Sarenac, D and Nsofini, J and Hincks, I and Arif, M and Clark, Charles W and Cory, DG and Huber, MG and Pushin, DA},
  journal={New journal of physics},
  volume={20},
  number={10},
  pages={103012},
  year={2018},
  publisher={IOP Publishing}
}

\end{document}